# Franz-Keldysh and Stark Effects in Two-Dimensional Metal Halide Perovskites

Kameron R. Hansen[1], C. Emma McClure[2], John S. Colton[2], and Luisa Whittaker-Brooks[1,*]

1. *Department of Chemistry, University of Utah, Salt Lake City, Utah 84112, United States*
2. *Department of Physics and Astronomy, Brigham Young University, Provo, Utah 84602, USA*

*Corresponding email: luisa.whittaker@utah.edu

**ABSTRACT**

As the field of metal halide perovskites (MHP) matures, state-of-the-art techniques to measure basic properties such as the band gap and exciton binding energy continue to produce inconsistent values. This issue is persistent even for 2D MHPs wherein the large separation between exciton and continuum states should make such measurements more straightforward. In this study, we revert to the established theory of a 2D Wannier exciton in a uniform electric field to analyze the electroabsorption response of an archetypal 2D MHP system, phenethylammonium lead iodide ($PEA_2PbI_4$). The high level of agreement between the electroabsorption simulation and measurement allows for a deepened understanding of the exciton's redshift according to the quadratic Stark effect and the continuum wavefunction leaking according to the Franz-Keldysh effect. We find the field-dependency of each of these effects to be rich with information, yielding measurements of the exciton's Bohr radius, transition dipole moment, polarizability, and reduced effective mass. Most importantly, the exciton binding energy is unambiguously determined with 2% uncertainty. The high precision of these new measurement methods opens the opportunity for future studies to accurately determine the influence of chemical and environmental factors on the optoelectronic properties of MHPs and thereby increase the tunability of this important class of materials.

I. **INTRODUCTION**

Excitons in 2D metal halide perovskites (2D MHPs) are stable at room temperature and are broadly viewed as analogous to those in inorganic quantum wells based on III-V semiconductors, such as GaAs-AlGaAs [1,2]. However, unlike III-V quantum wells which are grown with expensive epitaxial techniques to target defect-free crystals, 2D MHPs naturally crystallize into a heterostructure consisting of alternating organic-inorganic layers. Both the organic and inorganic components of the lattice impart distinct effects on the exciton. For example, the low-dielectric organic layer enhances the Coulomb interaction and causes an image charge effect, resulting in far greater exciton binding energies [3,4]. In a similar vein, phonons within the inorganic layer couple to the exciton states, resulting in a coexistence of multiple exciton-polarons with distinct phonon coherences [5,6]. The result is a unique tightly bound, highly polaronic, and moderately delocalized exciton that is becoming the focus of significant research efforts, both from fundamental and applied perspectives [7-10].

While the assembly of 2D MHPs into functional device structures has thus far been promising, these materials have yet to achieve the applied successes of their all-inorganic quantum well counterparts. To translate the unique photophysical properties of the excitons present in 2D MHPs into successful devices, a deeper understanding of the underlying absorption features as well as an understanding of how to tune their optoelectronic properties are needed. Unfortunately, the same lattice effects (e.g. dynamic disorder and phonon-coupling) which give excitons in 2D MHPs their unique character and defect-tolerant properties also result in spectral structures which are difficult to interpret. Nowhere is this more evident than with the difficulty in measuring exciton binding energies ($E_B$) for both 2D and 3D MHPs, a parameter of great importance for solar cell engineering as it determines the ratio of excitons to free carriers and for LED engineering as it strongly dictates electroluminescence quantum efficiency [11]. For example, taking the well-known 2D MHPs phenethylammonium lead iodide (PEA$_2$PbI$_4$) and butylammonium lead iodide (BA$_2$PbI$_4$) as case studies, state-of-the-art techniques for measuring $E_B$ in 2D MHPs, namely absorption, photoluminescence excitation, electroabsorption, and low- and high-field magnetoabsorption at liquid helium temperatures have thus far led to $E_B$ values ranging from 190 to 490 meV [12-17]. While some of this variance is caused by differences in the organic molecule, sample thickness, dielectric environment, and morphology, a significant portion of the variance originates from differing interpretations of band-edge absorption features and assumptions that are intrinsic to the measurement techniques. For theoretical and applied research directions within the MHP field to move forward, it is critical to develop high confidence, model-independent measurement methods that are precise and reproducible.

In pursuit of such measurements and a clear understanding of the optoelectronic properties in 2D MHPs, we are motivated to turn to the technique that was the most significant in advancing the fundamental understanding of III-V quantum wells, namely electroabsorption spectroscopy (EA). EA measures the difference in a material's absorption spectrum, $A$, with and without an applied field $F$, as follows:

$$\Delta A = A(F) - A(F = 0) \qquad (1)$$

In the 1980's, the seminal EA studies by Bastard, Mendez, and Miller [18-22], built off the theory of Dow and Blossey [23,24], not only demonstrated the anisotropy of the III-V quantum well electronic structure but also achieved a high level of agreement between theory and experiment [25] which led to measurements of the exciton's radius, binding energy, effective mass, as well as the coherence length of Bloch states [26]. However, in comparison to III-V quantum wells, thus far EA studies on 2D MHPs have been sparse and contradictory in their analysis method as well as their assignment of features within the EA spectrum [14,15,27,28].

This study seeks to bridge the gap between the EA features of 2D MHPs and our understanding of them by simulating the EA response according to the same theory that was successful for III-V quantum wells, namely that of a 2D Wannier exciton in a uniform field [23,29], but in a regime that accurately represents 2D MHPs (i.e. low-field and low-broadening). We find that this theoretical description of the EA response for 2D MHPs allows for a deeper understanding of the exciton's redshift according to the quadratic Stark effect as well as the continuum wavefunction leaking into the forbidden gap according to the Franz-Keldysh effect. Our simulated EA spectrum matches closely with our experimental measurement, and we find that all the discrepancies between the simulation and measurement can be understood in the context of charge-lattice interactions which are not accounted for in the 2D Wannier exciton Hamiltonian. The line shape and field dependence of the measured EA response allow for direct measurements (using one fit parameter) of the exciton's radius, polarizability, transition dipole moment, and reduced effective mass. In particular, the 2% uncertainty in our model-independent measurement of $E_B$ opens the door for future studies to precisely track chemical and environmental effects on $E_B$ and thereby add another degree of tunability to the engineering of efficient MHP devices.

## II. RESULTS AND DISCUSSION

### A. Sample Preparation

Pinhole-free, 2D MHP thin films with ~25 μm grains and a thickness of 80 – 120 nm were created by spin-coating a precursor solution containing a 2:1 ratio of phenethylammonium iodide (PEAI) and lead iodide (PbI$_2$) to form PEA$_2$PbI$_4$ on a substrate of interdigitated electrode fingers (see **Sec. III A** for fabrication details). The small spacing between opposing electrodes (45 μm) allowed for the generation of high applied fields (11 – 155 kV/cm) with moderate voltages (50 – 700 V) parallel to the substrate. As shown in **Fig. 1(a)**, the PEA$_2$PbI$_4$ crystal consists of alternating organic-inorganic nanolayers and thus is classified as a natural-forming quantum well material [30]. Given the high degree of anisotropy in the electronic structure of quantum wells, EA responses for 2D MHPs differ substantially when the applied field is parallel vs. perpendicular to the quantum well layers [31]. To determine the quantum well orientation, we collected grazing-incidence wide-angle X-ray scattering (GIWAXS) patterns for PEA$_2$PbI$_4$ films. The GIWAXS pattern in **Fig. 1(b)** shows c-axis diffraction peaks corresponding to the (00*l*) family of crystallographic planes, confirming that the 2D layers in PEA$_2$PbI$_4$ are predominantly stacked with the c-axis normal to the substrate [9]. To quantify the homogeneity of the grain orientations, we calculated the mosaicity factor (MF) of the off-c-axis diffraction of the (002) peak. The MF value falls on a linear scale from -1 to 1, corresponding to homogeneous grain orientation perpendicular (-1) and parallel (1) to the substrate [32]. The PEA$_2$PbI$_4$ thin film's MF value is 0.964 and therefore, we assume an ideal parallel case in our analysis of the EA response.

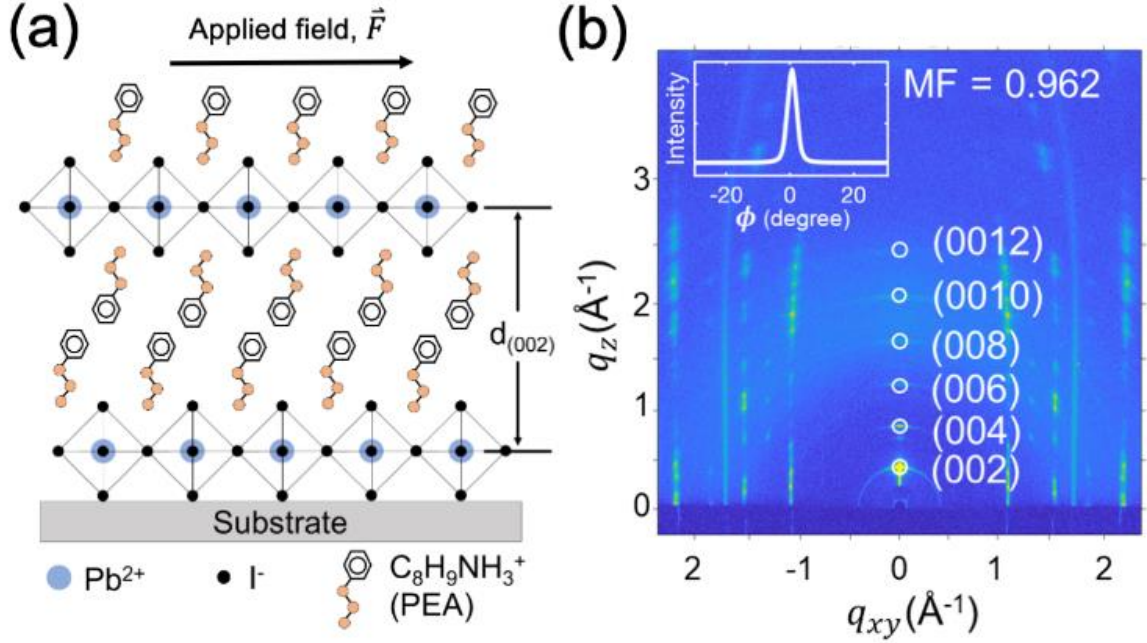

FIG. 1 (a) 2D MHP crystal structure based on a phenethylammonium lead iodide with general formula $A_2BX_4$, where in this case A = phenethylammonium (PEA$^+$), B = lead (Pb$^{2+}$), and X = iodide (I$^-$). (b) GIWAXS pattern showing c-axis diffraction peaks corresponding to the (00$l$) family of crystallographic planes. Converting these peaks to real space via Bragg's law, we find the $d_{002}$ spacing to be 1.629 ± 0.003 nm. The inset shows a phi-cut of the (002) diffraction peak which is used to calculate a mosaicity factor of 0.962, indicating the quantum wells are preferentially oriented parallel to the substrate.

### B. Simulation

In 2D MHPs, confinement occurs perpendicular to the quantum wells, and therefore, the in-plane EA response probed in this study is free of the quantum-confined Stark effect and quantum-confined Franz-Keldysh effect. Instead, excitons below the band gap respond to external fields via the quadratic Stark effect while continuum states at the band gap respond via the Franz-Keldysh effect [19]. The influence of these two effects on the EA response can be precisely accounted for using the theory of Dow and Lederman for a 2D Wannier exciton in a uniform electric field [29], based on three input parameters specific to the material: (1) the exciton's spectral linewidth $\Gamma$ (also known as the material's homogenous line broadening), (2) the exciton's reduced effective mass $m^*$, and (3) the material's effective dielectric value, $\varepsilon_{r,eff}$. For MHPs with highly frequency-dependent dielectric properties, the dielectric value of relevance here likely falls in the infrared range (~$10^{13}$ Hz) [33] and is denoted as an effective dielectric value $\varepsilon_{r,eff}$ which is measured by inputting the binding energy and mass into the Bohr model description of the exciton binding energy [33,34].

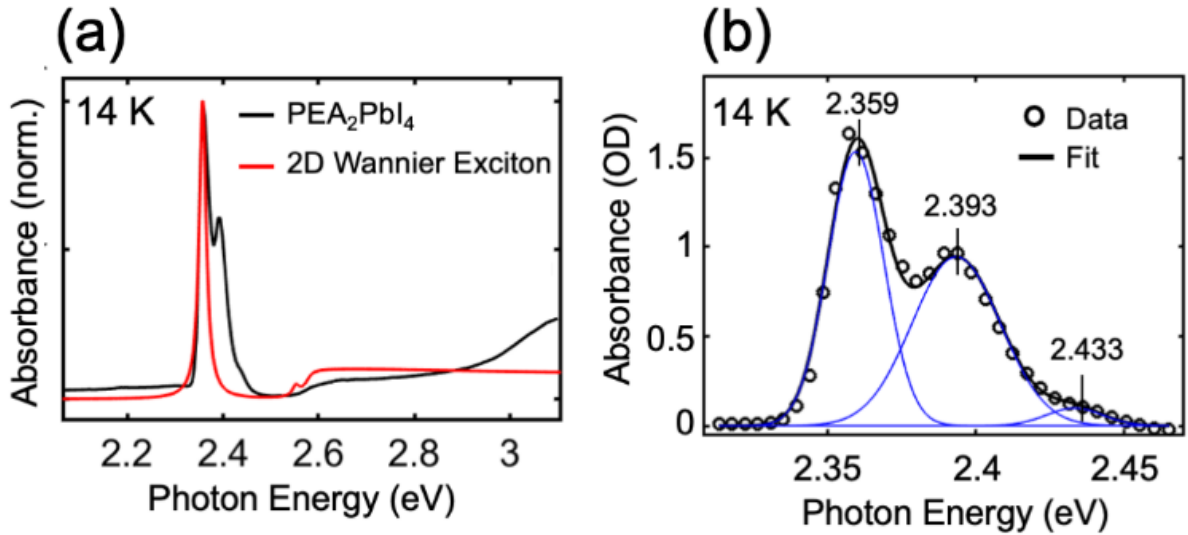

FIG. 2 (a) Absorption spectrum (black) of a PEA$_2$PbI$_4$ thin film compared to that of a 2D Wannier exciton (red). (b) Three Lorentzian fit profiles to the 1s exciton absorption peak and its two phonon sidebands at higher energies. The half-width-half-maximum of the peak at 2.359 eV is measured to be 11 meV and is taken as the material's homogeneous line broadening, $\Gamma$.

The homogenous line broadening Γ = 11 meV is found by fitting a Lorentzian profile to the exciton's absorption peak, as shown in **Fig. 2**. At first glance, there are several notable discrepancies between the absorption spectrum of PEA$_2$PbI$_4$ (black line in **Fig. 2(a)**) and that of an ideal 2D Wannier exciton (red line). First, the exciton absorption for PEA$_2$PbI$_4$ exhibits three peaks with 35 ± 5 meV spacing, as shown in **Fig. 2(b)**, which have been extensively studied [7,35] and recently have been shown to originate from excitonic coupling to distinct, coherent phonon modes within the [PbI$_6$]$^{4-}$ layer [5,6]. Second, the 2s exciton absorption (small peak near 2.57 eV) is not resolved for PEA$_2$PbI$_4$ which is a result of the image charge effect suppressing the oscillator strength of the 2s state, an effect that is unique in 2D MHPs [3]. Lastly, at above-gap energies, PEA$_2$PbI$_4$ and other 2D MHPs exhibit multiple absorption peaks that likely arise from an overlap of the exciton continuum of states with secondary conduction band minima, which band structure calculations consistently predict to be present in 2D MHPs [12,36].

Having obtained Γ, the EA spectrum can be simulated using reasonable estimates for $m^*$ and $\varepsilon_{r,eff}$ in 2D MHPs. These material parameters enter the theory through the so-called ionization field ($F_I$), or the field strength required to create a potential drop of one Rydberg ($R$) energy across the exciton's Bohr radius ($a_0$):

$$F_I = R/ea_0 \qquad (2)$$

where the Rydberg energy and Bohr radius have their standard definitions:

$$R = \frac{\hbar^2}{2a_0^2 m^*} \qquad (3)$$

$$a_0 = \frac{4\pi\hbar^2 \varepsilon_0 \varepsilon_{r,eff}}{e^2 m^*} \tag{4}$$

And it is the ratio between the field strength $F$ and the ionization field $F_I$,

$$f = F/F_I \tag{5}$$

that ultimately determines if the EA spectrum should be simulated in the high-field regime ($f > 1$) where exciton states are ionized, or in the low-field regime ($f < 1$) where ionization effects are negligible. The field strengths $F$ in this study range from 2.5 – 34.5 kV/cm [37] while we estimate the ionization field for 2D MHPs fall in the range of $10^2 - 10^4$ kV/cm assuming reasonable values for $\varepsilon_{r,eff}$ between 3 – 5 and a reduced effective mass $m^*$ between $0.050 - 0.221\, m_0$ [4,12]. Therefore, taking $F < F_I$, we simulate the EA spectrum using the theory of reference [29] for an ideal 2D Wannier exciton in the low-field and low-broadening regime ($f < 1$, $\Gamma \ll R$). We find a surprisingly high level of qualitative agreement between the measured EA response for PEA$_2$PbI$_4$ (**Fig. 3(a)**) and the simulated EA response (**Fig. 3(b)**). For details on the numerical simulation, see **Sec. III C**.

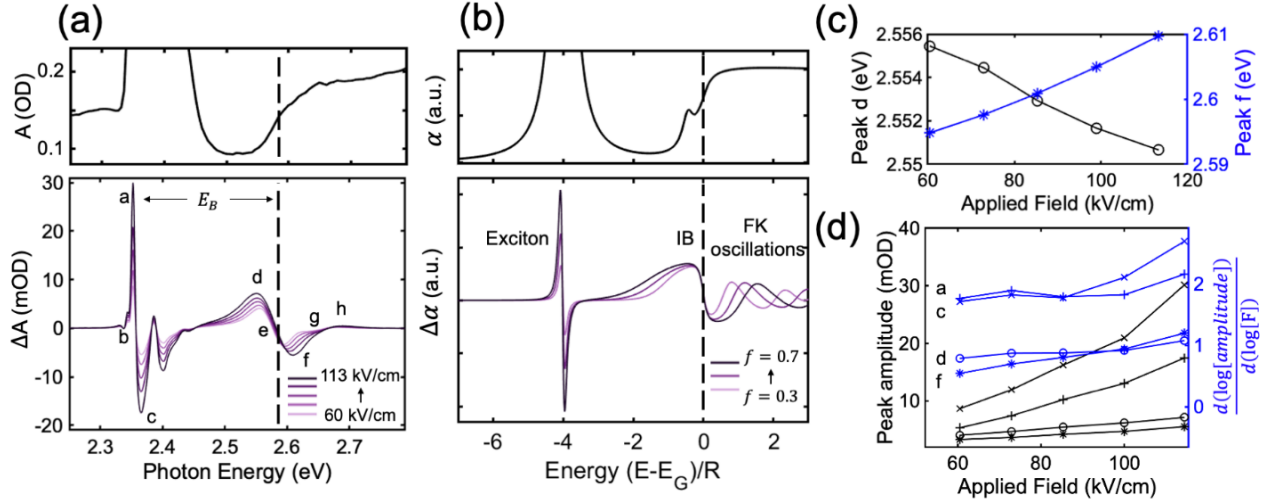

FIG. 3 (a) Absorption (top) and EA (bottom) spectra for PEA$_2$PbI$_4$ at 14 K in comparison with (b) the simulated absorption (top) and EA (bottom) spectra for a 2D Wannier exciton in the low-field, low-broadening regime. (c) The energy of 'd' and 'f' peaks as a function of applied field. The position of 'd' decreases in energy with increasing applied field while the position of 'f' increases as the continuum wavefunction leaks further into the forbidden gap. (d) Peak amplitudes as a function of applied field. The derivative of log(amplitude) with respect to log(F), shown on the right-hand y-axis, provides the power law of each curve and reveals that the exciton peaks 'a' and 'c' scale quadratically with the field in comparison to the IB Franz-Keldysh peaks 'd' and 'f' which scale linearly.

### C. $E_G$ and $E_{1s}$ assignments

As displayed in **Fig. 3(a,b)**, the two primary features of the simulated and measured EA response for PEA$_2$PbI$_4$ are: 1) a first-derivative feature due to the exciton's Stark shift (labeled a – c in **Fig. 3(a)**), and 2) a single oscillatory period at the interband (IB) absorption (labeled d – h). These features have been resolved in previous EA studies on 2D MHPs and while the exciton feature clearly originates from the Stark effect, the high-energy IB feature has received contradictory interpretations leading to miscalculations of $E_G$ and $E_B$ [14,15,27,28]. We find that the established theory of Dow and Lederman [29] predicts the IB feature, as shown in the simulated spectrum in **Fig. 3(b)**. Furthermore, the energetic position exactly at $E_G$ reveals that the origin of the IB feature lies in the continuum wavefunction leaking into the forbidden gap (discussed further in **Sec. II F**). The field dependence of the simulated IB feature is precisely matched in the experimental EA response. With increasing field strength, the wavefunction of PEA$_2$PbI$_4$ is expected to (and does) leak further into the gap. This field dependence is plotted in **Fig. 3(c)** where peak 'd' redshifts with increasing field and peak 'f' blueshifts, indicating increased transfer of oscillator strength from above the gap to below the gap. The crossover point where the change in the oscillator strength switches from positive to negative (zero-crossing labeled 'e') is taken as a high confidence measurement of the one-electron band gap energy of $E_G = 2.579 \pm 0.004$ eV for PEA$_2$PbI$_4$.

The assignments of the Stark and IB features are further corroborated by the field dependence of the PEA$_2$PbI$_4$ EA amplitudes. We plot the amplitude of exciton peaks ('a' and 'c') and IB peaks ('d' and 'f') as a function of applied field in **Fig. 3(d)** and find the exciton features 'a' and 'c' scale with $F^2$, which is expected for third order ($\chi^{(3)}$) nonlinear optical processes such as the Stark effect [38]. Meanwhile, the Franz-Keldysh effect is only a $\chi^{(3)}$ response when the field-perturbation on the continuum states is small compared to the homogenous broadening $\Gamma$ [39]. This is not the case for the EA response for PEA$_2$PbI$_4$, as we find that the IB features scale linearly with $F$. The redshift of the exciton peak due to the quadratic Stark effect results in a sharp, first derivative EA line shape with zero-crossing at point 'b' marking the 1s exciton resonant energy of $E_{1s} = 2.357 \pm 0.002$. The exciton binding energy, $E_B = E_G - E_{1s}$ is then found to be $222 \pm 4$ meV where uncertainty primarily arises from the slight variation in zero-crossing points for each field strength. Despite the fact that our simulations are founded in the hydrogenic model, we emphasize that this measurement of $E_B$ is model-independent since the Stark shift and wavefunction leaking phenomenon occur at critical points $E_{1s}$ and $E_G$ in the PEA$_2$PbI$_4$ density of states.

Table I. Exciton binding energies averaged from multiple thin films

|  | $E_B$ |
|---|---|
| PEA$_2$PbI$_4$ (< 45 K) | 223 ± 3 |
| PEA$_2$PbI$_4$ (300 K) | 222 ± 10 |
| BA$_2$PbI$_4$ (< 45 K) | 250 ± 4 |
| BA$_2$PbI$_4$ (300 K) | 205 ± 7 |

Repeating this EA experiment, we find that the variation of $E_B$ for different samples is also low. Three PEA$_2$PbI$_4$ thin films and two BA$_2$PbI$_4$ thin films (where BA = butylammonium) were tested with slightly different thicknesses and processing conditions, yielding $E_B$ values of {222 ± 4, 222 ± 6, 225 ± 5} meV for PEA$_2$PbI$_4$ and {251 ± 2, 248 ± 8} meV for BA$_2$PbI$_4$ (averaged values provided in **Table I**). We choose to study BA$_2$PbI$_4$ because, similar to PEA$_2$PbI$_4$, it is among the most commonly studied 2D MHP systems. However, unlike PEA$_2$PbI$_4$ (and similar to nearly all other 2D MHP compositions), its absorption spectrum lacks a step-like feature at the band gap [40,41]. Despite lacking this feature in the absorption spectrum, we find that the IB feature is present in the EA spectra for BA$_2$PbI$_4$ (see **Appendix A**), demonstrating that our measurement method for $E_B$ is robust and likely can be applied to any MHP system with sufficient separation between exciton and continuum states ($E_B \gg \Gamma$). We find the difference in organic cation (PEA vs BA) has essentially no effect on the EA signal- the two primary features, exciton and IB, are robustly observe with the same field dependence across all tested thin films. This is expected given that previous studies have established the organic cation primarily only affects the electronic states by modulating the dielectric environment [40,42]. The full data set of EA spectra and measurements of $E_{1s}$, $E_G$, and $E_B$ are provided in **Appendix A**.

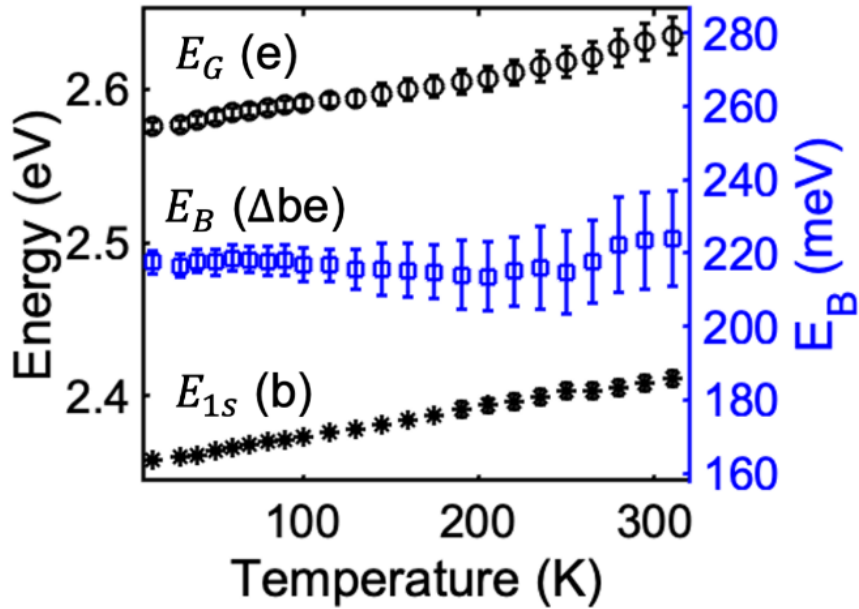

FIG. 4 Energetic positions of $E_G$ given by zero-crossing 'e', $E_{1s}$ given by zero-crossing 'b', and $E_B$ given by the difference between $E_G - E_{1s}$. The uncertainty in $E_{1s}$ remains low compared to the uncertainty in $E_G$ which increases with temperature given the decrease in the amplitude of the IB feature.

Moving to higher temperatures, the broadening $\Gamma$ increases and this measurement of $E_B$ becomes less precise, as shown in **Fig. 4** for PEA$_2$PbI$_4$. The $E_{1s}$ and $E_G$ levels blueshift with increasing temperature consistent with the unique anti-Varshni effect expected for MHP systems [43] while $E_B$ remains relatively constant. The uncertainty in both $E_G$ and $E_B$, however, rise monotonically with temperature as the zero-crossing point 'e' become a less trustworthy as a marker for $E_G$ due to greater homogenous and inhomogeneous broadening mechanisms. It should also be noted that for all samples, we observed the exciton's EA line shape transition from first derivative to second derivative with increasing temperature, as detailed in **Appendix B**. Therefore for $T \gtrsim 150$ K, the zero-crossing point 'b' can no longer be taken as $E_{1s}$. Rather, $E_{1s}$ is more accurately measured by the resonance peak in the absorption spectrum with an uncertainty determined by the peak's intrinsic broadening. The high- and low-temperature values of $E_B$ for PEA$_2$PbI$_4$ and BA$_2$PbI$_4$ are summarized in **Table I**.

### D. Discrepancies between theory and experiment

The most noticeable discrepancy between the simulated and measured spectra lies in the additional oscillation near 2.4 eV the PEA$_2$PbI$_4$ spectrum. By plotting the EA alongside the first derivative of the absorption spectrum (**Fig. 5**), it is evident that the two spectra have nearly identical line shapes across the entire exciton range 2.32 – 2.44 eV indicating that all the excitonic absorption peaks are uniformly redshifted by the applied field [14,44]. Thus, we assign this additional oscillation to the quadratic Stark shift of the phonon sideband. Another notable discrepancy is that sinusoidal-like oscillations at high energies are present in the simulated EA spectrum, but not the measured spectrum. These are Franz-Keldysh oscillations and can persist over a wide spectral range in high-quality GaAs-based multiple quantum wells [26], but are not expected when conduction band Bloch states have a short coherence length. The absence here demonstrates the magnitude of phase-disrupting perturbations in the charge carrier's energy landscape and/or the frequency of carrier-lattice scattering events, which are expected for the highly disordered, highly ionic PEA$_2$PbI$_4$ MHP lattice [45]. Thus, we find that the discrepancies in the theory and experiment are consistent with charge-lattice effects which are well documented for 2D MHPs, but which are not accounted for in the Hamiltonian of the 2D Wannier exciton. Meanwhile, the two main EA features, those being the exciton and the IB features, are well captured in the simulation. Having obtained a high-confidence measurement of $E_B$, we can proceed to analyze the field dependency of these features according to Stark and Franz-Keldysh theories.

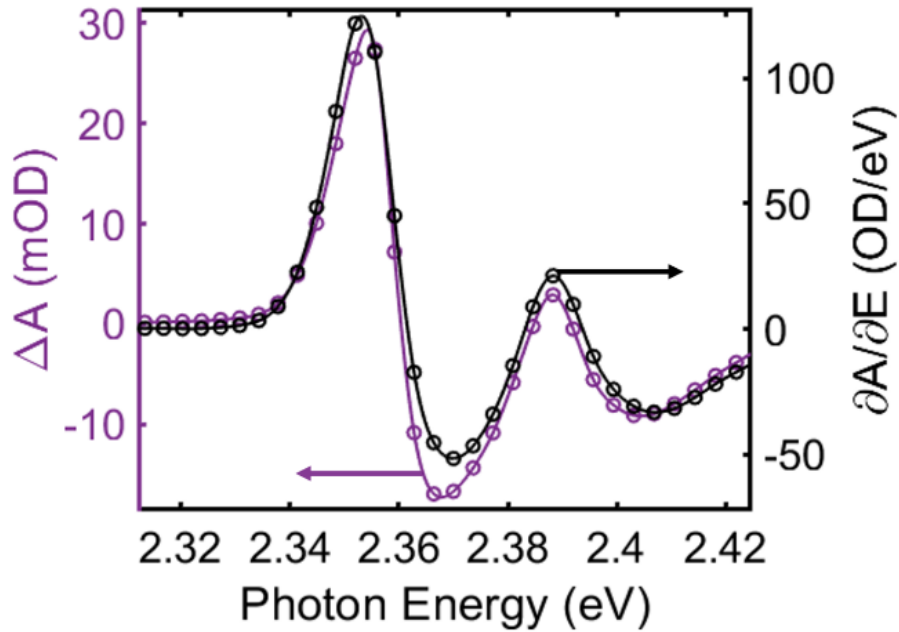

FIG. 5 The EA spectrum (left, purple) in comparison to the first derivative of the absorption spectrum (right, black) in the exciton region for PEA$_2$PbI$_4$. The oscillation at 2.39 eV is caused by a Stark shift of the phonon sideband.

### E. Stark shift of 1s exciton

Unlike exciton states in III-V quantum wells, the large energetic separation between $E_{1s}$ and continuum states places the excitons in 2D MHPs within a regime where ionization effects are negligible and where a perturbative approach to analyzing the exciton Stark shift is well-justified. Thus, the magnitude of the linear Stark shift ($\Delta E_L$) and quadradic Stark shift ($\Delta E_Q$) of the 1s exciton of PEA$_2$PbI$_4$ can be independently measured by fitting the EA response to first and second derivatives of the unperturbed absorbance [46] as per the following equation:

$$\Delta A = \frac{\partial A}{\partial E}\Delta E_Q + \frac{1}{2}\frac{\partial^2 A}{\partial E^2}(\Delta E_L)^2 \tag{6}$$

**Fig. 6** shows the measured $\Delta E_Q$ for two PEA$_2$PbI$_4$ samples, along with the theoretical values predicted for two- and three-dimensional Wannier excitons, as given by **Eq. (7)** and **Eq. (8)** [29,44].

$$\frac{\Delta E_{Q,3D}}{E_B} = -\frac{9}{8}\left(\frac{ea_0 F}{E_B}\right)^2 \tag{7}$$

$$\frac{\Delta E_{Q,2D}}{E_B} = -\frac{7}{12}\left(\frac{ea_0 F}{E_B}\right)^2 \tag{8}$$

Here, $E_B$ is the measured binding energy (223 meV), $a_0$ is the exciton's 3D Bohr radius as defined in **Eq. (4)**, and $F$ is the field strength, i.e. the applied field screened by the dielectric value at the frequency of the modulating voltage, $\varepsilon_r$ (983 Hz). We measured this dielectric value to be 4.5 (see **Appendix C**). The smaller Stark shift in the 2D limit is due to the symmetry of the charge distribution in two dimensions reducing the exciton's polarizability. The relation in **Eq. (6)** holds when field ionization is negligible, while **Eq. (7)** and **Eq. (8)** are valid so long as $\Delta E$ is small compared to $R$. Both requirements are well satisfied for 2D MHPs at our experimental field strengths, and therefore, we use $\Delta E_Q$ to determine the exciton's Bohr radius. Requiring the Stark shift to fall between the 3D and 2D limits, we adjust the single fit parameter $a_0$ in **Eq. (7)** and **Eq.**

(8) to find a lower bound of 1.9 nm and a lower bound of 2.8 nm for PEA$_2$PbI$_4$ (shown in **Fig. 6** for $a_0$ = 2.35 nm). In its initial state, the exciton's true radius likely falls between the 3D Bohr radius ($a_0$ = 2.35 nm) and 2D Bohr radius ($a_0/2$ = 1.18 nm) limits meaning it is delocalized over a length of ~2 – 4 [PbI$_6$]$^{4-}$ octahedra parallel to the quantum well layer. However, more detailed theoretical models of MHPs must be developed to determine how subsequent polaron formation affects the radius on longer timescales [6]. Although this is the first experimental measurement of $a_0$ in MHPs that we are aware of, our window of 1.18 – 2.35 nm for the exciton's radius captures the RMS wavefunction extension, measured by Dyksik et al. as $\sqrt{\langle r^2 \rangle}$ = 1.3 nm for PEA$_2$PbI$_4$ [17].

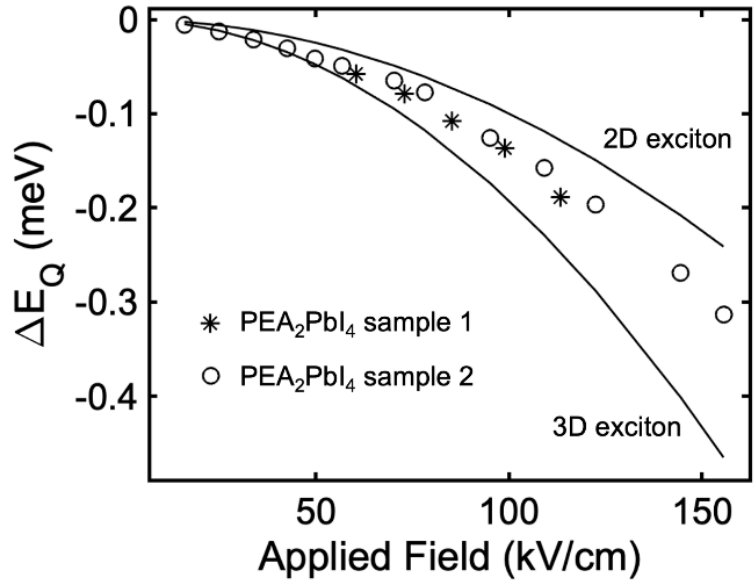

FIG. 6 In-plane quadratic Stark shift of the 1s exciton for PEA$_2$PbI$_4$, compared to limits expected for 2D and 3D Wannier excitons, set by Eq. (7) and (8), respectively.

For each field strength, the magnitude of the measured $\Delta E_L$ and $\Delta E_Q$ values can be straightforwardly converted to the exciton's transition dipole moment $\mu_{ge}$ and polarizability $\alpha_{ge}$ according to the definitions of the quadratic and linear Stark effects [44]:

$$\Delta E_L = -\mu_{ge} F \qquad (9)$$

$$\Delta E_Q = -\frac{1}{2}\alpha_{ge} F^2 \qquad (10)$$

where $\mu_{ge}$ is the change in permanent dipole moment and $\alpha_{ge}$ is the change in polarizability upon a transition from the ground state to the excited state $g \to e$. We find our measurement of $\mu_{ge} = 13 \pm 4$ D for our PEA$_2$PbI$_4$ thin film to be in close agreement with the value of 11 D measured in a recent optical Stark effect study by Proppe et al. [47]. The magnitude of the quadratic Stark shift yields a large polarizability of $80000 \pm 10300$ Å$^3$ expected for a delocalized Wannier exciton. This polarizability value is about ten times greater than excitons in molecular matrices [44] and three times greater than in inorganic quantum dots [48]. A summary of these measurements is provided in **Table II**.

Table II: Exciton Properties for PEA$_2$PbI$_4$ at 15 K.

|  | PEA$_2$PbI$_4$ |
|---|---|
| $a_0$ (nm) | $2.35 \pm 0.55$ nm |
| $\mu_{ge}$ (D) | $13 \pm 4$ |
| $\alpha_{ge}$ (Å$^3$) | $80000 \pm 10300$ |
| $m^*$ (m$_0$) | $0.09 \pm 0.024$ |

### F. Franz Keldysh Effect

The physical origin of the IB feature is rendered clear in **Fig. 7(a)** by comparing the zero-field band-edge absorption strength $\alpha(E, f = 0)$ to that of the field-shifted spectra, $\alpha(E, f > 0)$. The maximum 'd' and minimum 'f', which are large in the EA response observed for all 2D MHPs even at room temperature, arise from the wavefunction leaking into the forbidden gap. Subsequent Franz-Keldysh oscillations at higher energies are less clear in the experimental EA spectra—only the first peak (labeled 'h') is resolved at low temperatures. However as discussed previously, these oscillations dampen rapidly and even the observed peak 'h' is extremely small. The clearest example of it is shown in **Fig. 7(b)** for $BA_2PbI_4$.

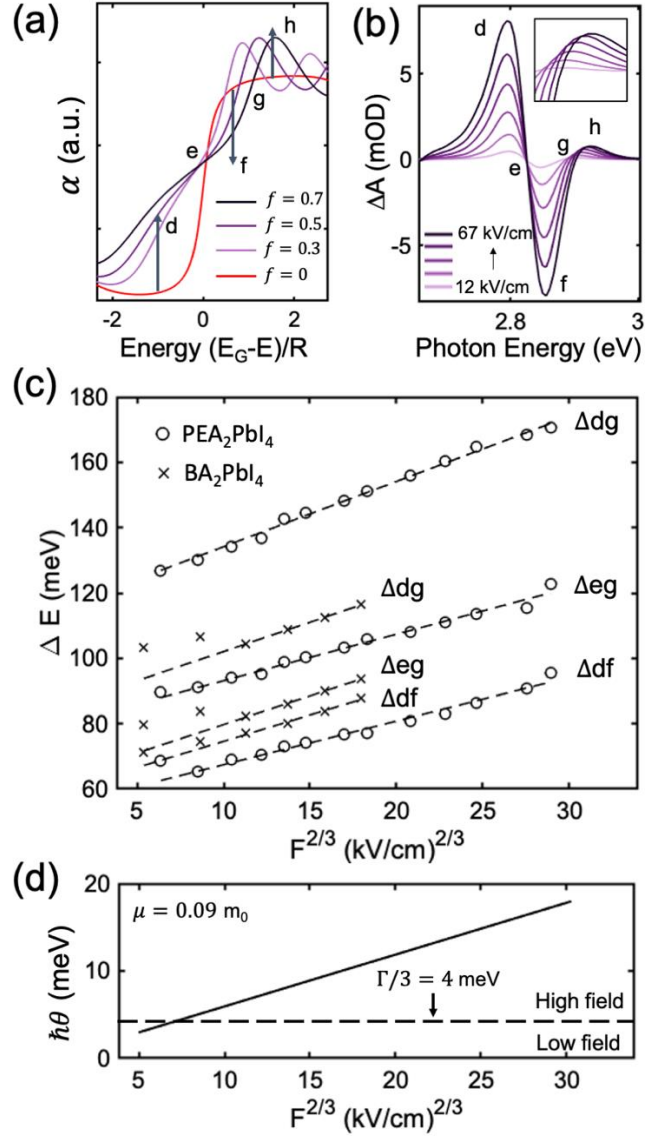

FIG. 7 (a) Band edge absorption with increasing field simulated for a 2D Wannier exciton. (b) IB EA feature and subsequent Franz-Keldysh oscillation resolved clearly for a $BA_2PbI_4$ sample. (c) Field broadening of the 'd' - 'h' EA features measured by the energetic separation between critical points in the EA spectrum and plotted as a function of the applied field $F^{2/3}$. Only data points well above the high-field Franz-Keldysh cutoff (F > 30 kV/cm) are included in the fit. (d) The electro-optic energy as a function of the applied field $F^{2/3}$. When $\hbar\theta$ is greater than 4 meV, the EA features should be analyzed according to high-field Franz-Keldysh theory.

The IB feature d – f and Franz-Keldysh oscillation g – h can be intuitively understood as the difference between exciting into Bloch vs Airy conduction band states, where the latter is the solution to an unbound electron in a uniform electric field. For a sufficiently high field and small mass, the oscillations in the EA spectrum take the form of an Airy function whose argument is the electron's kinetic energy scaled by the electro-optic energy, $\hbar\theta$, as follows [49,50]:

$$Ai\left(-\frac{E-E_G}{\hbar\theta}\right) \qquad (11)$$

$$\hbar\theta = (\hbar^2 e^2 F^2/2m^*)^{1/3} \qquad (12)$$

where $F$ is the field strength and $m^*$ is the reduced effective mass. In the high-field regime (defined as $\hbar\theta > \Gamma/3$), the amplitudes of Franz-Keldysh oscillations are expected to scale according to $\sqrt{\hbar\theta} \propto F^{1/3}$, while positions should broaden according to $\hbar\theta \propto F^{2/3}$ according to the Airy function behavior [51]. Indeed, at high applied fields, we observe that the spectral broadening between relative points d – h has a consistent slope proportional to $F^{2/3}$, as shown in **Fig. 7(c)**. As detailed in **Appendix D**, we use the slope to solve for the exciton reduced effective mass using one fit parameter, $m^*$. We find $m^* = 0.09 \pm 0.024$ $m_0$ for PEA$_2$PbI$_4$, and $m^* = 0.08 \pm 0.02$ $m_0$ for BA$_2$PbI$_4$, where the majority of the uncertainty arises from sample-to-sample variation in the rate of field broadening. Given that the band structure for these 2D MHP compositions originate from the same [PbI$_6$]$^{4-}$ octahedra layer, we expect the masses to be the same. We find that these values are consistent with measurements in 3D MHPs (0.12 $m_0$ by Ziffer et. al [52] and 0.108 $m_0$ by Miyata et al. [11]) as well as Dyksik et al.'s high-field magnetoabsorption measurement of 0.091 $m_0$ [16] and 0.087 $m_0$ [17] for PEA$_2$PbI$_4$, but does not agree with Blancon et al.'s low-field magnetoabsorption measurement of 0.221 $m_0$ for BA$_2$PbI$_4$ [12].

In **Fig. 7(d)**, we plot the resulting electro-optic energy using the mass for PEA$_2$PbI$_4$ and find that the electro-optic energy surpasses the homogenous broadening ($\Gamma/3$) near an applied field of 18 kV/cm (or 6.9 (kV/cm)$^{2/3}$). Indeed, this is concomitant with the field strength at which the field-broadening in **Fig. 7(c)** acquires a consistent slope proportional to F$^{2/3}$, thus supporting the measurements of $m^*$ and $\Gamma$. For lower quality samples or samples at higher temperatures, the high-field Franz-Keldysh limit will increase proportional to the broadening $\Gamma$.

In conclusion, we find that the dielectric and quantum confinement effects which enhance $E_B$ by an order of magnitude place 2D MHPs in a unique category where the exciton enters the high-field regime at an ionization field of $F_I$ = 944 kV/cm (calculated according to **Eq. (2)**), but the continuum enters the high-field Franz-Keldysh regime at a much lower 18 kV/cm. This is due to the fact the high-field cutoff for the exciton depends on its radius and binding energy, while the continuum depends on the broadening and the exciton mass. We have shown that the field dependence in the EA line shape and amplitude can be used to measure the exciton's binding energy, radius, transition dipole moment, polarizability, and reduced effective mass. These values place meaningful bounds on future theoretical models for 2D MHP systems. The modest in-plane radius (1.18 – 2.35 nm) and polarizability (80000 ± 10300 Å$^3$) measured herein demonstrate that excitons in 2D MHPs are quasi-2D Wannier excitons delocalized over 2 – 4 [PbI$_6$]$^{4-}$ octahedra parallel to the quantum wells, resulting in a 2:1 to 4:1 asymmetry ratio. Despite this less-than-ideal 2D nature, we find that the qualitative similarities between the EA response for 2D MHPs and that of an ideal 2D Wannier exciton are strong. The simulation of the 2D Wannier exciton, based on the established theory of Dow and Lederman, deepens our understanding of the IB feature in the EA response of 2D MHPs and leads to a high-confidence, model-independent measurements of $E_G$ and $E_B$. This method for measuring $E_B$ in MHP systems opens the door for future studies to precisely track chemical and environmental effects on $E_B$ and thereby acquire more control of this important variable in MHPs as they are assembled into functional devices.

## III. METHODS

### A. Film fabrication and characterization

We synthesized millimeter-sized PEA$_2$PbI$_4$ and BA$_2$PbI$_4$ single crystals by dissolving PbO (0.536 g, 2.4 mmol) and the organic amine (phenethylamine (756 μL, 6mmol) or butylamine (593 μL, 6mmol)) in excess of hydroiodic acid (5 mL) stabilized with a small addition of hypophosphorous acid (75 μL) [53]. All chemicals were purchased from Sigma Aldrich. The solution was stirred under heat (~70 °C) for ~5 minutes until the solutes dissolved, and then removed from the hotplate resulting in the precipitation of a bright orange solid. The crystals were then filtered with diethyl ether and dried under a vacuum at 60 °C for two hours. The filtered and dried single crystals were brought into a glovebox (~0.1 ppm O$_2$) and dissolved in a 4:1 DMF:DMSO mixture to target a 0.07 – 0.1 M solution and then stirred for 2 hours at 60 °C. We pipetted the resulting precursor solution onto a 1.2 × 1.2 cm quartz substrate with interdigitated gold electrodes (**Fig. 8**), spin-cast at 4500 rpm for 20 seconds, then annealed the film for 30 minutes at 100 °C. The low molarity and high spin casting speed produced thinner films which was advantageous for EA in order to both resolve the exciton's peak at low transmission and avoid saturating the detector at wavelengths of large transmission (peak optical density ≈ 1 – 2).

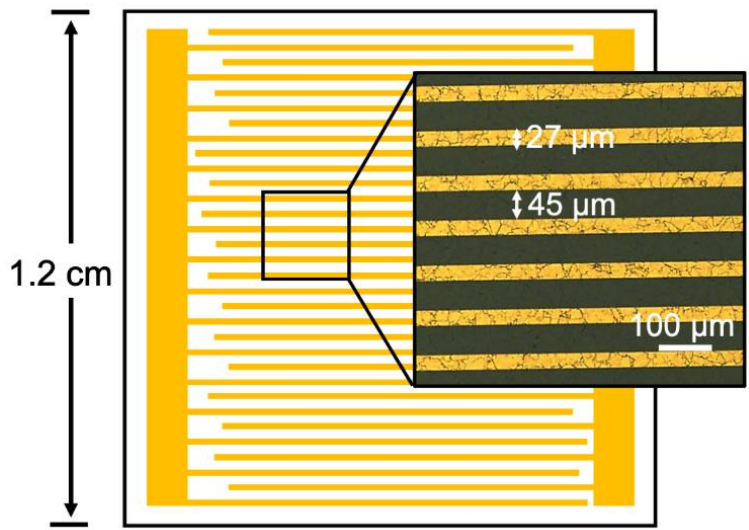

FIG. 8 Substrate used for EA and dielectric spectroscopies. Gold electrodes were deposited via photolithography onto a quartz substrate with 45 μm spacing between opposing electrodes. The inset is an optical micrograph of a $PEA_2PbI_4$ thin film on the interdigitated fingers.

For structural characterization, GIWAXS images were acquired on beamline 11-3 at the Stanford Synchrotron Radiation Lightsource (SSRL). A 12.7 keV X-ray beam was angled at 0.2° with respect to the substrate which is below the substrate's critical angle and above the film's critical angle. The diffraction pattern was collected on a 3072 × 3072 pixel CCD array positioned 315 mm from the sample. All images were background, polarization, and absorption corrected using a GIXSGUI software [54]. Distortion from the latter two effects was negligible. The MF values were calculated according to the procedure outlined in reference [32].

## B. Electroabsorption spectroscopy

Each sample consisted of a MHP thin film deposited on a photolithographically patterned array of interdigitated gold electrodes with a spacing $d = 45$ µm which allowed for the generation of high electric fields $> 10$ kV/cm with modest voltages ~50 – 700 V, similar to that used in previous electroabsorption studies [55]. The sample was mounted in a cryostat for transmission of sample $T$, transmission of substrate $T_0$, and electro-transmission of sample $\Delta T$. Incandescent light from a Xe lamp was dispersed through a Digikröm 0.25 m monochromator with 1 nm spectral resolution, focused on the sample, and detected by a Thorlabs SM1PD2A photodiode detector. The RMS value of the photodiode's signal was demodulated with a Stanford Research Systems SR810 lock-in amplifier operating in current mode. For $T$ and $T_0$, the signal was detected at the first harmonic of the mechanical chopper's frequency (330 Hz), whereas the $\Delta T$ signal was lock-in detected at either the first or second harmonic of the modulating signal, $V_{AC}(t)$, depending on which Fourier component of the squared signal $V_{AC}^2(t)$ was largest. Various $V_{AC}(t)$ waveforms were tested and found to yield consistent EA line shapes and amplitudes after instrumental correction factors were accounted for. The reported absorbance was calculated as $A = \log_{10}(T_0/T)$. The EA response is then given by:

$$\Delta A = \log_{10}\left(\frac{T_0}{T_F}\right) - \log_{10}\left(\frac{T_0}{T}\right) \quad (13)$$

where $T_F$ represents the field-perturbed transmission spectrum. Substituting $T_F = \Delta T + T$, **Eq. (13)** becomes:

$$\Delta A = -\log_{10}\left(1 + \gamma \frac{\Delta T}{T}\right) \quad (14)$$

where an instrumental correction factor $\gamma$ has been introduced, which is the ratio of the RMS value of the mechanical chopper signal's $1\omega$ Fourier coefficient and the RMS value of the $V_{AC}^2(t)$ signal's $1\omega$ (or $2\omega$) Fourier coefficient [56].

### C. Numerical Simulation

The numerical simulations for a 2D Wannier exciton in a uniform electric field were run on the frisco1 cluster at the Center for High Performance Computing at the University of Utah using in-house code written for Matlab R2020b. The numerical procedure we implemented is based on references [23] and [57], and we summarize it here for the two-dimensional case.

We begin with the Elliot formula for direct interband transitions, which relates a semiconductor's absorption coefficient $\alpha$ to the electron-hole wavefunction $\psi_n(0)$ as follows [58,59]:

$$\alpha(E) = \frac{2\pi |d_{cv}|^2 \omega}{n_b \varepsilon_0 c} \sum_n |\psi_n(0)|^2 \delta(E_n - \hbar\omega) \tag{15}$$

where $d_{cv}$ is the dipole matrix element, $\hbar\omega$ is the photon energy, $n_b$ is the background refractive index, $\varepsilon_0$ is the permittivity of free space, and $c$ is the speed of light. To obtain the absorption coefficient, we must first solve for $\psi_n(0)$ which in our case corresponds to the wavefunction of a 2D Wannier exciton in a uniform field. In dimensionless units, this Schrödinger equation takes the following form:

$$\left(-\nabla_{x,y}^2 - \frac{2}{r} + fx\right)\psi_n(x,y) = E_n \psi_n(x,y) \tag{16}$$

where $\psi_n(x, y)$ is the wave function for the electron-hole relative motion, $E_n$ is $\hbar\omega - E_G$, $f$ is a dimensionless field defined in **Eq. (5)**, and the label "n" represents a quantum state with energy $E$, azimuthal quantum number $m$, and a separation parameter (or "parabolic eigenvalue") $t_n$. To make **Eq. (16)** tractable, a transformation to parabolic coordinates is needed, as follows:

$$\xi = r + z, \ \zeta = r - z, \ x = \frac{1}{2}(\xi - \zeta) \tag{17}$$

A separation of variables assumption is made in parabolic coordinates, and we seek solutions of the following form:

$$\psi_n(r) = \frac{\chi_1(\xi)\chi_2(\zeta)}{(\xi\zeta)^{1/4}} \tag{18}$$

Inserting this ansatz into **Eq. (16)** results in the following two quasi-Schrödinger equations:

$$\chi_1(\xi)'' + \left(\frac{1-m^2}{4\xi^2} + \frac{t_n}{\xi} + \frac{E}{4} - \frac{f\xi}{8}\right)\chi_1(\xi) = 0 \tag{19}$$

$$\chi_2(\zeta)'' + \left(\frac{1-m^2}{4\zeta^2} + \frac{1-t_n}{\zeta} + \frac{E}{4} - \frac{f\zeta}{8}\right)\chi_2(\zeta) = 0 \tag{20}$$

Here, $t_n$ represents the separation parameter which must be numerically solved for, $J$ is a parameter that sets the Coulomb interaction, and $m = \pm 0.5$ corresponds to the azimuthal number for direct interband transitions in two-dimensions. Thus, the problem of solving **Eq. (16)** is reduced to solving for $\chi_1(\xi)$ and $\chi_2(\zeta)$. For given value of $f$ and $E$, the numerical procedure begins by setting $J = 1$, $m = 0.5$, and using an array of initial guesses for $t_n$ spanning a range of approximately 0.1 to 4. The potential in **Eq. (19)** is positive as $\xi \to \infty$ and therefore an exponentially decaying $\chi_1$ is the first boundary condition with an asymptotic form that we determine using the WKB approximation. For each $t_n$, the Numerov method [60] is then used to integrate $\chi_1(\xi)$ inward until its first maximum is reached. At the origin, both $\chi_1(\xi)$ and $\chi_2(\xi)$ can be expanded as a power series, as follows:

$$\chi_{1-UN} = \xi^{(m+1)/2}\left(1 - \frac{t_n\xi}{m+1} + O(\xi^2)\right) \tag{21}$$

$$\chi_{2-UN} = \xi^{(m+1)/2}\left(1 - \frac{(1-t_n)\xi}{m+1} + O(\xi^2)\right) \tag{22}$$

This power series approximation $\chi_{1-UN}$ is similarly integrated outward from the origin until the previously determined maximum point is reached, at which point, the two halves of $\chi_1$ are rescaled to make the amplitude continuous at the joining point. The Cooley method [61] is then implemented to compare the slopes at the joining point and determine a correction factor:

$$DE = \frac{\chi'_{1,out} - \chi'_{2,out}}{\int_0^\infty (\chi(\xi))^2 d\xi} \tag{23}$$

This correction factor is calculated for each $t_n$ within the array and a smoothing spline is fit to the function $DE(t_n)$ in order to accurately determine its zero-crossings. The function $DE(t_n)$ oscillates around zero with first, second, third, etc. zero-crossing nodes corresponding to the $n = 1, 2, 3,$ etc. parabolic eigenvalues $t_n$. These values are then used as initial values of $t_{n,i}$ and further iterative integration ($t_{n,i+1} = t_{n,i} - DE$) is carried out convergence is obtained ($DE < 0.02 * 2^{floor(i/10)}$). This converged value of $t_n$ is considered final and used to find $\chi_2(\zeta)$. We insert $t_n$ into **Eq. (22)** and integrate $\chi_{2-UN}(\zeta)$ outward from the origin according to **Eq. (20)**. The $\chi_2(\zeta)$ function is expected to converge to its unbound asymptotic form, which Dow and Redfield found to be [23]:

$$\chi_2(\zeta) \approx \frac{A}{(\zeta/2 + E/f)^{1/4}} \sin\left(\frac{2}{3} f^{\frac{1}{2}} \left(\frac{E_n}{f} + \frac{1}{2}\zeta\right)^{3/2} + \delta\right) \quad (24)$$

where $A$ is a normalization constant that must be solved for. At large $\zeta$, **Eq. (23)** is fit (with 0.5% accuracy) to the integrated $\chi_{2-UN}(\zeta)$ and the normalization constant $A$ is thereby determined. At this point, all the information necessary to calculate the field-shifted absorption coefficient $\alpha_F$ has been determined. As shown in references [23] and [59], $\alpha_F$ depends on $A$ and $\chi_{1-UN}(\xi)$ as follows:

$$\alpha_F(E) \cong \alpha_0 \frac{E_G}{R} \sum_n \left(A^2 \pi^2 f^{\frac{1}{2}} \int_0^\infty \frac{\chi^2_{1-UN}(\xi)}{\xi} d\xi\right)^{-1} \quad (25)$$

where $\alpha_0$ is the 2D Wannier exciton absorption coefficient that depends on $d_{cv}$, $n_b$, and the Bohr radius $a_0$ [59]. In our calculations, we set the prefactor $\alpha_0 E_G/R$ equal to one and only compare the simulated vs measured EA line shape, not the amplitude. The summation in **Eq. (25)** is carried out for all $n$ that contribute more than 0.1% to the total absorption coefficient and we find typically $n_{max} = 1, 2$ or 3. The $m = -0.5$ case is computed and the two absorption coefficients are summed together. Repeating this procedure for a range of $E$ produces the field-shifted absorption spectrum $\alpha_F(E)$. Analytical solutions to **Eq. (19)** and **Eq. (20)** exist for the case of $f = 0$, which results in the following zero-field absorption spectrum [29]:

$$\alpha(E) \cong C \left( \frac{1}{\pi} \sum_n^\infty \frac{1}{\left(n-\frac{1}{2}\right)^3} \delta\left(E + \left(n-\frac{1}{2}\right)^{-2}\right) + \frac{1}{2\pi} \frac{\Theta(E)}{1+e^{-2\pi/\sqrt{E}}} \right) \qquad (26)$$

where $C$ is a normalization factor and $\Theta(E)$ is the Heaviside step function. The $\alpha_F(E)$ and $\alpha(E)$ spectra are then integrated and $C$ is adjusted to make the spectral weight equal. Next, the two spectra $\alpha_F(E)$ and $\alpha(E)$ are numerically convolved with a Lorenztian function, $L(E)$ with a half-width-half-maximum (HWHM) of $\Gamma$ and subtracted to produce the simulated EA spectrum $\Delta\alpha$:

$$\Delta\alpha = \alpha_F(E) * L(E) - \alpha(E) * L(E) \qquad (27)$$

We use an $L(E)$ with $\Gamma = 0.15$, which is our measurement of the PEA$_2$PbI$_4$ homogenous line broadening $\Gamma = 11$ meV in units of $R$. We note that $R$ can be reasonably determined multiple ways. The standard conversion for 2D systems is $E_B = 4R$, however this is inexact for 2D MHPs since the excitons are not fully confined to a 2D plane and they additionally experience an $E_B$-enhancing image charge effect. A more accurate value for $R$ can be determined by inputting our measurements of $a_0$ and $m^*$ into **Eq. (3)**, which results in $R = 76$ meV. If instead we use our measurement of $m^*$ and $\varepsilon_r = 4.3$ (see **Appendix C**), this results in $R = 69$ meV. We choose to average the two and use $R = 72.5$ meV for our conversion of $\Gamma$.

Using our measurement of $F_I = 944$ kV/cm, the field strengths in this study correspond to an excitonic dimensionless field ranging from 0.004 to 0.06. However, these values of $f$ correspond to a theoretical exciton peak width from ~$10^{-1152} R$ to ~$10^{-74} R$, only a slight increase from the zero-field case of a Dirac-delta function [19]. Therefore, the field-shifted absorption of 2D MHPs in the excitonic range cannot be accurately simulated even with quadruple precision in $E$. However, we find that simulating in a slightly higher field range $0.3 < f < 0.7$, as shown in the main text, elucidates the origin of the IB feature and therefore is sufficient for the purposes of this study.

The IB absorption, on the other hand, can be simulated at lower fields. We find the linear field dependence of the IB feature's amplitude to be well captured by the simulation at low fields $0.1 < f < 0.2$, as shown in **Fig. 9**. We choose this dimensionless field range which is ~3 times larger than the exciton's dimensionless field since the IB absorption is purely determined by the electro-optic energy relative to the homogenous broadening, and therefore is not affected the by $E_B$-enhancing image charge effect which places the exciton in the extremely low-field range.

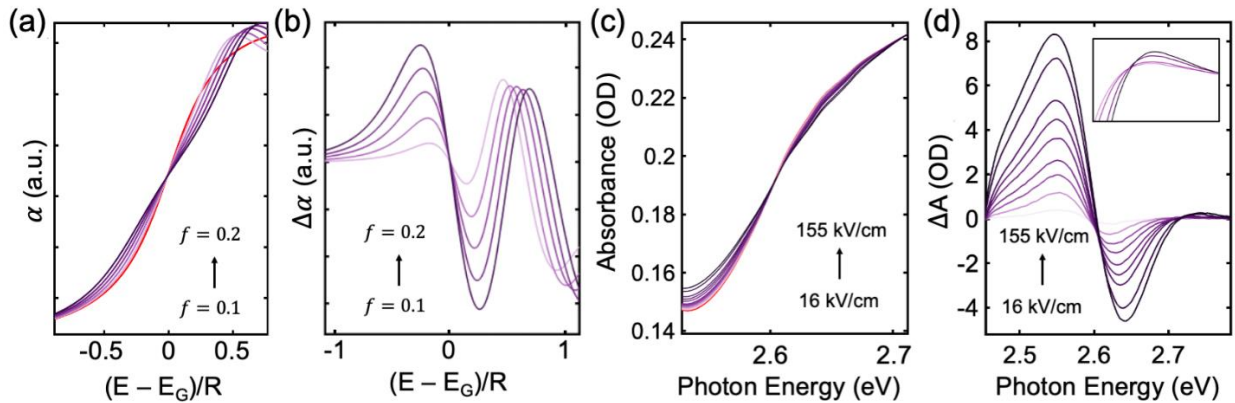

FIG. 9 (a) Simulated absorption and (b) EA in the IB range for small dimensionless fields 0.1 – 0.2. (c) The measured field-shifted absorption and (d) EA in the IB range for PEA$_2$PbI$_4$.

**APPENDIX A: EA DATASET**

In **Table III**, we provide the full dataset of $E_{1s}, E_G$, and $E_B$ measurements for five 2D MHP samples at low and high temperatures. The BA$_2$PbI$_4$ composition undergoes a phase transition near ~250 K [13,15] while PEA$_2$PbI$_4$ does not. The IB feature's zero-crossing is taken as $E_G$, while exciton's zero-crossing is taken as $E_{1s}$ only for low temperatures. Above ~150 K, $E_{1s}$ is more accurately determined as the maximum of the exciton's absorption peak. The ~20 meV sample-to-sample variation in $E_{1s}$ and $E_G$ is likely primarily caused by differences in the sample's temperature, given that the difference $E_G - E_{1s}$ is remarkably consistent. Room temperature EA was not measured for PEA$_2$PbI$_4$ sample 2.

Table III. $E_{1s}, E_G$, and $E_B$ measurements all 2D MHP samples

| | | | |
|---|---|---|---|
| Low Temperature (≤50K) | | | |
| | $E_{1s}$ (eV) | $E_G$ (eV) | $E_B$ (meV) |
| PEA$_2$PbI$_4$ sample 1 | 2.357 ± 0.002 | 2.579 ± 0.004 | 222 ± 4 |
| PEA$_2$PbI$_4$ sample 2 | 2.378 ± 0.002 | 2.603 ± 0.005 | 225 ± 5 |
| PEA$_2$PbI$_4$ sample 3 | 2.384 ± 0.004 | 2.606 ± 0.004 | 222 ± 6 |
| BA$_2$PbI$_4$ sample 1 | 2.551 ± 0.0015 | 2.802 ± 0.0015 | 251 ± 2 |
| BA$_2$PbI$_4$ sample 2 | 2.578 ± 0.008 | 2.826 ± 0.002 | 248 ± 8 |
| High Temperature (~300K) | | | |
| | $E_{1s}$ (eV) | $E_G$ (eV) | $E_B$ (meV) |
| PEA$_2$PbI$_4$ sample 1 | 2.411 ± 0.01 | 2.635 ± 0.012 | 224 ± 16 |
| PEA$_2$PbI$_4$ sample 3 | 2.404 ± 0.01 | 2.624 ± 0.0035 | 220 ± 11 |
| BA$_2$PbI$_4$ sample 1 | 2.422 ± 0.01 | 2.631 ± 0.002 | 209 ± 10 |
| BA$_2$PbI$_4$ sample 2 | 2.455 ± 0.01 | 2.656 ± 0.003 | 201 ± 10 |

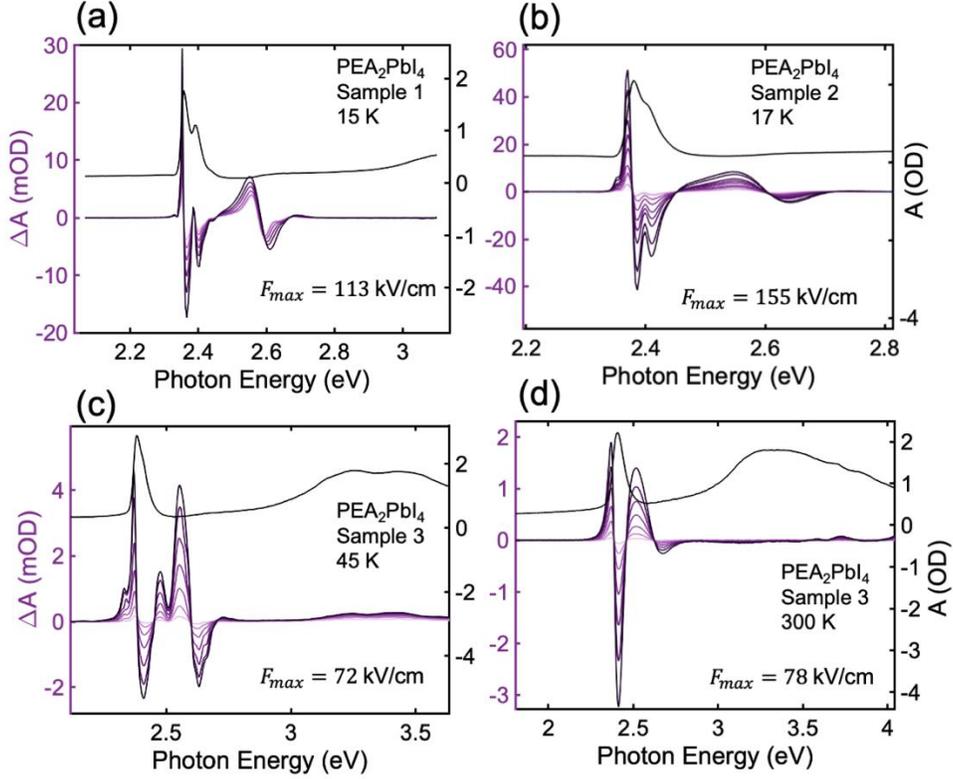

FIG. 10 Low and high temperature EA spectra of three PEA$_2$PbI$_4$ thin films. A voltage-series for the 300 K spectrum was only measured for PEA$_2$PbI$_4$ sample #3.

The field-dependent EA spectra corresponding to the measurements in **Table III** are displayed in **Fig. 10** for PEA$_2$PbI$_4$ and **Fig. 11** for BA$_2$PbI$_4$. Both materials share the same EA response, a first-derivative line shape in the exciton range that scales proportional to $F^2$ and an IB feature across the band gap that scales proportional to $F$. This difference in field dependence results in a smaller

IB feature, in comparison to the exciton, for the EA spectrum that are extended to high fields (PEA$_2$PbI$_4$ sample 1 in **Fig. 10(a)** and sample 2 in **Fig. 10(b)**). Due to a coexistence of high- and low-temperature crystal phases in the BA$_2$PbI$_4$ samples [13,15], two exciton peaks (Stark shifts) are observed in the absorption (EA) spectrum at low temperature (**Fig. 11(a) and Fig. 11(c)**). Lastly, we note that for PEA$_2$PbI$_4$ sample 3, BA$_2$PbI$_4$ sample 1, and BA$_2$PbI$_4$ sample 2, there was a difference in lamp power between the substrate transmission and the sample transmission measurements, and therefore, the absorption spectra for these samples should be viewed as approximations. For this reason, the quantitative analysis of these samples was limited to the EA spectra, which depends only on the electrotransmission and transmission of the sample. For PEA$_2$PbI$_4$ sample 1, we present the temperature-dependent absorption and EA in **Appendix B**.

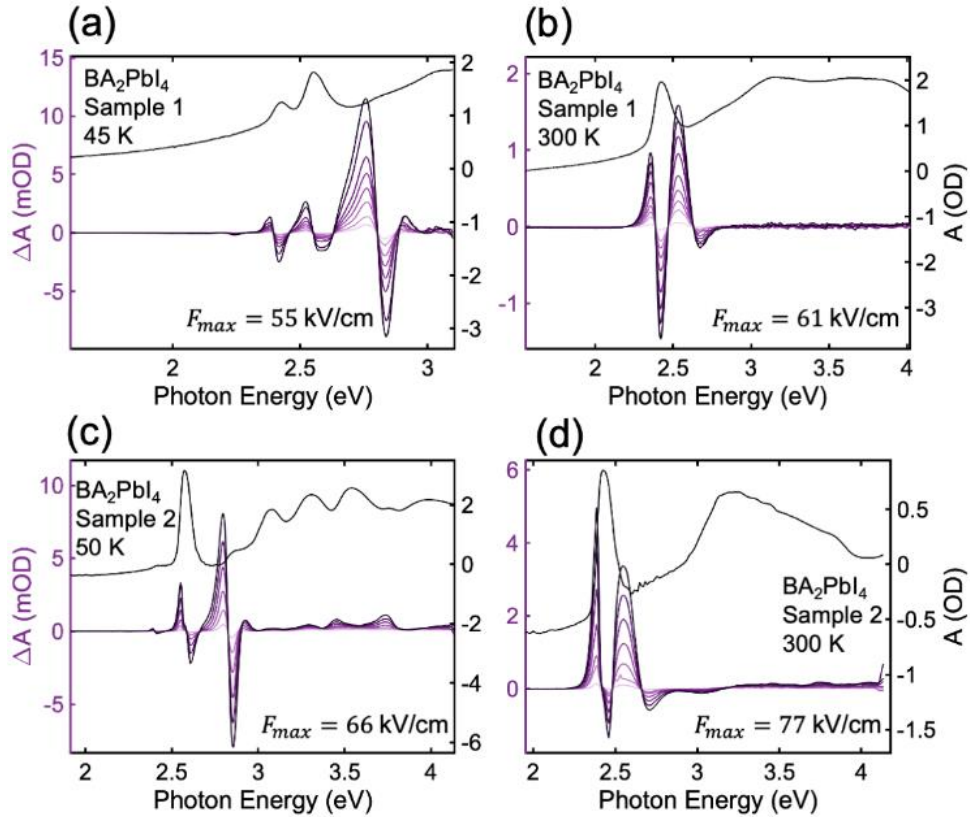

FIG. 11 Low and high temperature EA spectra for two $BA_2PbI_4$ thin films. The difference between the exciton and IB features was used to measure the exciton binding energy and the rate of the field broadening in the Franz-Keldysh oscillation was used to measure the excitonic reduced effective mass for $BA_2PbI_4$.

## APPENDIX B: TEMPERATURE-DEPENDENT EA

We observe that the exciton's EA line shape transition from first to second derivative moving from low to high temperature for all the 2D MHPs tested in this study, as shown in **Fig. 12**. For this reason, the exciton's zero-crossing point in the EA spectrum is not a valid measurement of $E_{1s}$ for temperatures $\gtrsim$ 150 K, but rather, is more accurately determined by the free exciton absorption peak. To the best of our knowledge, strong temperature dependence in the EA line shape has previously only been observed for solvated dye molecules [62]. We suspect the transition is caused by an increased dipole moment which would result in a second-derivative line shape according to the linear Stark effect, as is clear by **Eq. (6)** and **Eq. (9)**. Disorder has been shown to induce large dipole moments in molecular systems [63]. A similar effect may be present here, but with the disorder arising from the thermal motion of the lattice rather than static morphological disorder in a molecular matrix, however, further testing of this hypothesis is beyond the scope of this study.

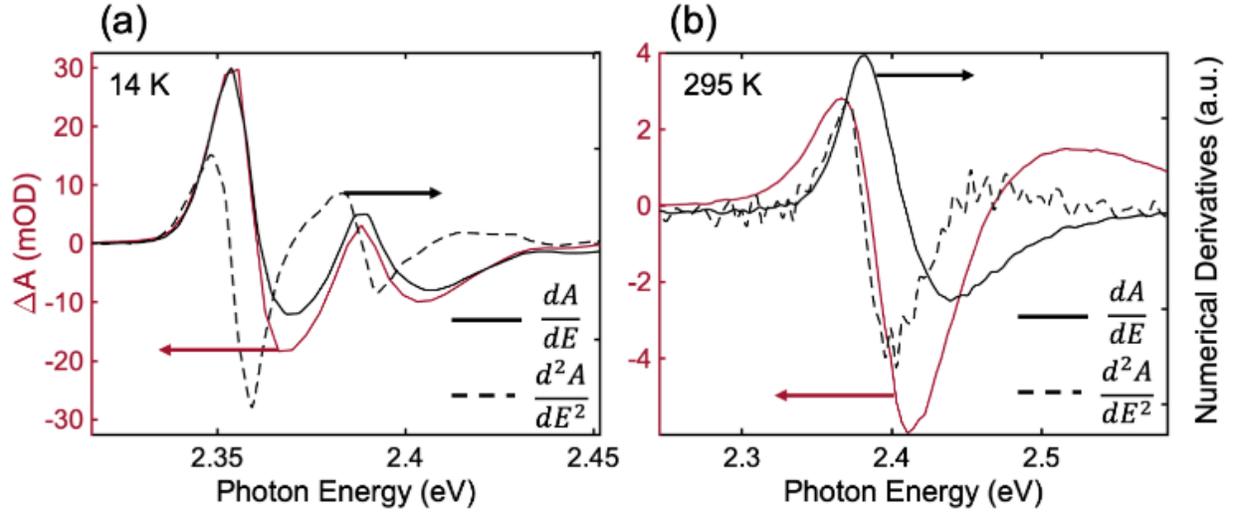

FIG. 12 (a) The low temperature EA spectrum (red) for PEA$_2$PbI$_4$ compared to first (solid black) and second (dotted black) numerical derivatives of the absorbance spectrum. The EA line shape closely matches the first derivative. For first derivative line shapes, the peak of the underlying feature is given by the position of the zero crossing. In contrast, (b) the room temperature EA spectrum more closely resembles the second derivative of the absorbance and the zero-crossing point is an inaccurate measurement of the peak position.

Absorption and EA spectra were collected at intervals of 10 – 15 K moving from low temperature to room temperature, as shown in **Fig. 13**, for an applied field of 113 kV/cm. We measured the homogenous broadening at room temperature to be $\Gamma = 55$ meV, thus, an applied field greater than 165 kV/cm would be needed to observe high-field Franz-Keldysh behavior ($\hbar\theta > \Gamma/3$).

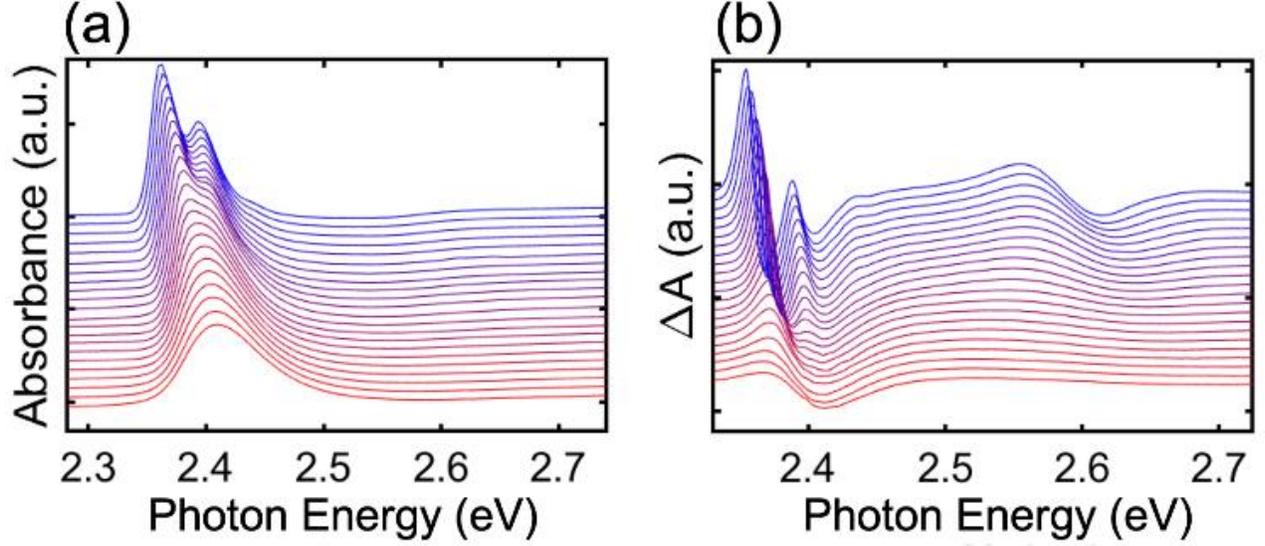

FIG. 13 (a) Absorbance and (b) EA spectra of PEA$_2$PbI$_4$ sample 1 from 30 K (blue) to 310 K (red) using an applied field of 113 kV/cm.

## APPENDIX C: MEASUREMENT OF DIELECTRIC VALUE

For the measurements of $a_0$, $\mu_{ge}$, $\alpha_{ge}$, and $m^*$ it is necessary to know the applied field strength experienced by the exciton, which depends on the RMS amplitude of the modulating square wave $V_{AC}$, the distance between opposing interdigitated electrode fingers $d$, and the dielectric constant at the frequency of the modulating voltage, $\varepsilon_r$(983 Hz):

$$F = \frac{V_{AC}}{d\,\varepsilon_r(983\text{ Hz})} \tag{28}$$

To obtain, $\varepsilon_r$(983 Hz), we performed dielectric spectroscopy using two geometries. First, a 1.54 mm-thick pressed pellet of PEA$_2$PbI$_4$ was placed in a parallel plate geometry and the capacitance as a function of frequency was collected using a Hewlett Packard 4192A Impedance Analyzer. The capacitance values were converted to the real part of the complex permittivity according to the air-gap method in reference [64]. These results are plotted as circles in **Fig. 14(a)**. Second, the capacitance of ~2 μm-thick $PEA_2PbI_4$ films on interdigitated fingers was measured and converted to the dielectric value according to equations in reference [65] which we modified to fit the thin film geometry [66]. As shown in **Fig. 14(a)**, the two methods agree at high frequencies and $\varepsilon_r(13\ MHz)$ ranges from 4.1 to 4.5. However, at low frequencies the dielectric values rise exponentially at different rates, likely caused by ion migration effects common in MHPs [67]. However, we find this exponential rise to be highly dependent on the film thickness. As shown in **Fig. 14(b)**, the normalized capacitance between 100 and $10^5$ Hz falls by two orders of magnitude for the 2.61 μm-thick film, whereas the capacitance of the 100 nm-thick film merely decreases by a factor of 0.81. The thicknesses of the films used for EA measurements ranged from 80 – 100 nm. Unfortunately, however, the capacitance of the 100 nm-thick film cannot be converted to $\varepsilon_r$ because the film's thickness is comparable to the electrode thickness and therefore violates the model used to relate the capacitance of the interdigitated electrodes to the dielectric constant [65]. Therefore, in order to obtain the 983 Hz dielectric value in the 100 nm-thick film, we scale the measured high-frequency value of $\varepsilon_r(13\ MHz) = 4.3$ by the 5% capacitive increase between 13 MHz and 983 Hz observed for the 100 nm-thick film, resulting in a value of $\varepsilon_r(983\ Hz) = 4.5$. This value agrees with estimates in the literature [4,12]. Our conclusion that ion migration effects are small in the 100 nm-thick film was corroborated by frequency-dependent EA. In ~100 nm-thick films, the exciton's EA magnitude (which scales $\propto F^2$) was found to be frequency-independent in the 1000 – 2000 Hz range but decreased slightly moving from 1000 Hz to 500 Hz.

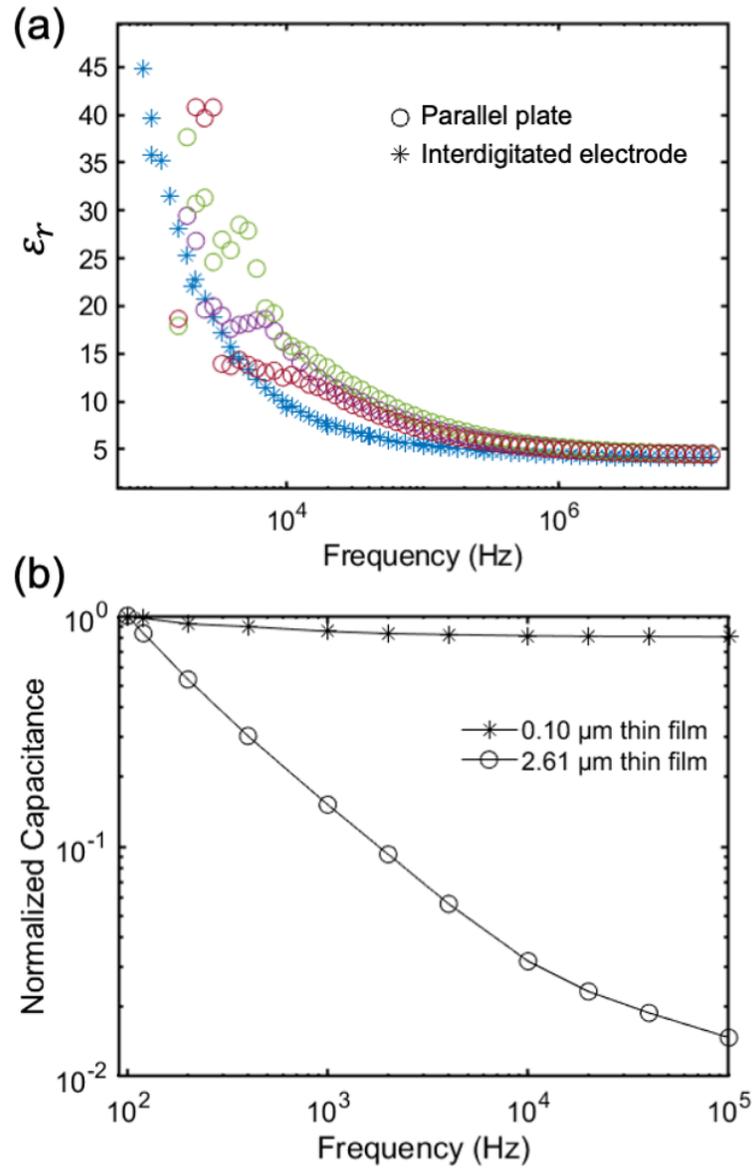

FIG. 14 (a) The real part of the PEA$_2$PbI$_4$ permittivity as a function of frequency measured with various methods: a 2.61 μm-thick film on interdigitated fingers (blue star), and a 1.54 mm-thick pressed pellet in an air-gap parallel plate geometry using a 1.28 mm air-gap (green circle), a 1.35 mm air-gap (green circle), and a 1.5 mm air-gap (red circle). (b) The normalized capacitance as a function of frequency for a thin (0.10 μm) and thick (2.61 μm) PEA$_2$PbI$_4$ film spin cast onto interdigitated electrodes.

# APPENDIX D: MEASUREMENT OF REDUCED EFFECTIVE MASS

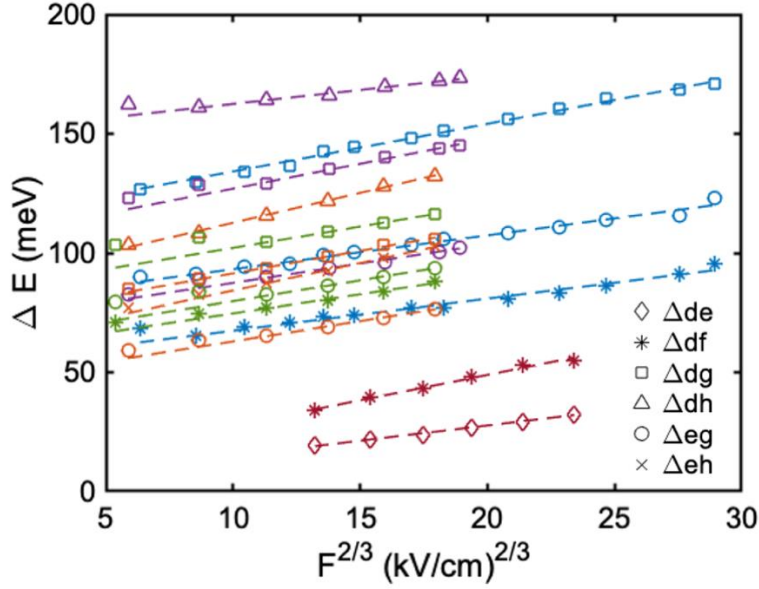

FIG. 15 Broadening of all EA features that were used to measure the reduced effective mass for $PEA_2PbI_4$ and $BA_2PbI_4$. The different colors represent broadening from the 5 samples tested (maroon, blue, purple = $PEA_2PbI_4$ samples 1, 2, 3; orange, green = $BA_2PbI_4$ samples 1, 2). Sample-to-sample variation between common critical points is quite large. For example, the magnitude of Δdh (triangle symbol) differs by 55 meV between $BA_2PbI_4$ sample 1 and $PEA_2PbI_4$ sample 3; however, we find the slopes to be sufficiently consistent to measure the exciton's reduced effective mass with $1\sigma$ variance of 27%.

In the high-field limit, Franz-Keldysh features broadening according to the argument of the Airy function in **Eq. (11)**. Point 'e' in the EA spectrum corresponds to zero while the first peak 'h' occurs at the Airy function's first root, 1.01879. Thus, the spacing between zero-crossing 'e' and the first positive peak 'h' in the EA spectrum depends on the electro-optic energy as follows:

$$\frac{h-E_G}{\hbar\theta} - \frac{e-E_G}{\hbar\theta} = 1.01879 \tag{29}$$

Solving for Δeh as a function of $F^{2/3}$ results in:

$$\Delta eh = 1.01879 \left(\frac{\hbar^2 e^2}{2m^*}\right)^{1/3} F^{2/3} \tag{30}$$

The mass is then determined from the slope of Δeh vs $F^{2/3}$. The field-broadening of points other than Δeh can also be used to determine the mass, provided the correct numerical factor is used in place of the Airy function's first root, 1.01879. We determined these numerical factors, relative to Δeh, from their spacing in the simulated 2D Wannier EA spectrum. These numerical factors are given in **Table IV**. To calculate the mass, we included broadening between all Franz-Keldysh features wherein the peak/zero-crossing positions were clearly resolved. The mean and standard deviation of these slopes (shown in **Fig. 15**) were computed to yield the mass and standard error reported in the main text.

Table IV. Dimensionless spacing between points in the Franz-Keldysh EA feature

|        | Numerical factor |
|--------|------------------|
| Δeh    | 1.01879          |
| Δdh    | 1.37 Δeh         |
| Δdg    | 1.10 Δeh         |
| Δeg    | 0.70 Δeh         |
| Δdf    | 0.65 Δeh         |

**APPENDIX E: PHOTOLUMINESCENCE**

The high and low-temperature photoluminescence (PL) for $PEA_2PbI_4$ and $BA_2PbI_4$ is shown in **Fig. 16**. To collect PL, thin films were mounted in a cryostat and excited using a diode laser ($\lambda$ = 400 nm, P = 0.25 mW). The emission from the sample surface was focused into a spectrometer (JY Horiba Triax 550) and recorded on a CCD array (JY Horiba Synapse). In addition to the free exciton emission, the $PEA_2PbI_4$ PL spectrum has a broad defect emission which is consistent with previous studies [8]. In contrast, the $BA_2PbI_4$ low-temperature PL has two narrow peaks which we attribute to the free exciton emission in the high and low-temperature crystal phases [15]. We measured a 50 meV Stokes shift between absorption and emission for $PEA_2PbI_4$ and 60 meV Stokes shift for $BA_2PbI_4$, which we find to be in agreement with other studies [15,68,69].

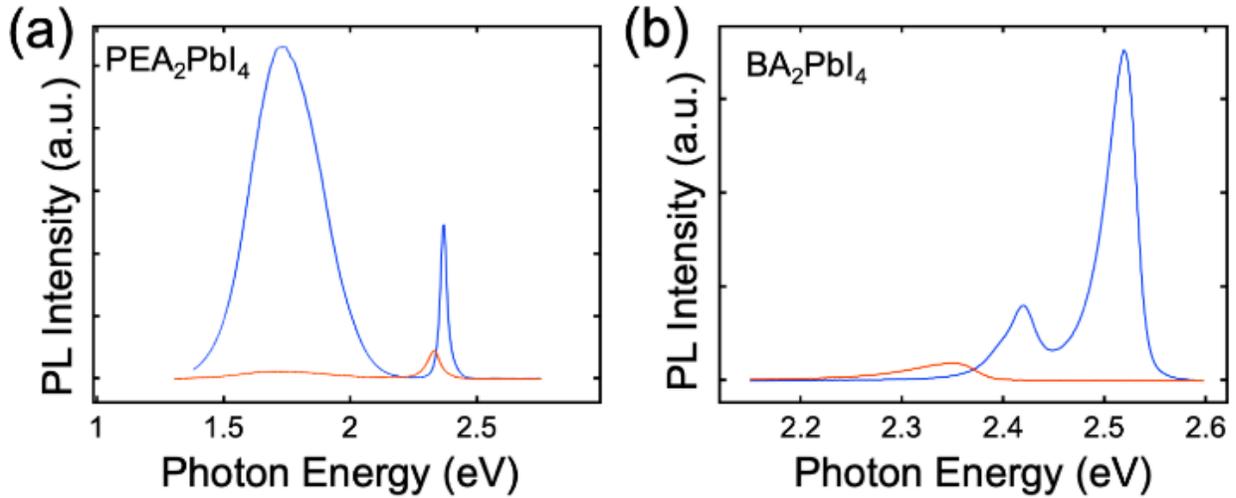

FIG. 16 (a) Photoluminescence of $PEA_2PbI_4$ at 15 K (blue) and 300 K (red) with the free exciton emitting at 2.33 and 2.37 eV, respectively. The broad low-energy peak has been extensively characterized in other studies as defect-state emission (b) Photoluminescence of $BA_2PbI_4$ at 45 K (blue) and 300 K (red) emitting at 2.52 and 2.36 eV, respectively.

**ACKNOWLEGMENTS**

This work was supported by the U.S. Department of Energy, Office of Basic Energy Sciences, Division of Materials Sciences and Engineering under Award no. DE-SC0019041. KH would like to acknowledge support from the National Science Foundation thru a Graduate Research Fellowship (Grant No. 1747505). Use of the Stanford Synchrotron Radiation Lightsource, SLAC National Accelerator Laboratory, is supported by the US DOE, Office of Science, Office of Basic Energy Sciences under contract DE-AC02-76SF00515. LWB would also like to acknowledge the Sloan Foundation through an Alfred P. Sloan Research Fellowship in Chemistry. We would also like to thank Dr. Mark Ziffer and Dr. XiYao Zhang for helpful advice on the numerical simulations.